\documentclass[aps,pra,onecolumn,groupedaddress,showpacs]{revtex4}

\usepackage{amssymb,amsmath}
\usepackage{graphicx}

\newcommand{\om}{\omega}

\newcommand{\pa}{\partial}

\begin{document}

\title{Breather solutions of the modified Benjamin-Bona-Mahony equation}

\author{G. T. Adamashvili}
\affiliation{Technical University of Georgia, Kostava str.77, Tbilisi, 0179, Georgia.\\ email: $guram_{-}adamashvili@ymail.com.$ }

\begin{abstract}
New two-component vector breather solution of the modified Benjamin-Bona-Mahony (MBBM) equation is  considered. Using the generalized perturbation reduction  method  the MBBM equation is reduced to the coupled nonlinear Schr\"odinger  equations for auxiliary functions. Explicit analytical expressions for the profile and parameters of the vector breather oscillating with the sum and difference of the frequencies and wavenumbers are presented.
The two-component vector breather and single-component scalar breather of the MBBM equation is compared.
\end{abstract}


\maketitle

\section{Introduction}

A solitary wave is a localized highly stable nonlinear wave that maintains its profile when it propagates at a constant speed through a medium.
The existence of solitary waves  in the form of a soliton and its various modifications, including breather, vector soliton, and others,   is  one of the most striking demonstrations of the nonlinear properties of the medium in which they are propagating.
Such waves are encountered  in a wide variety of different field of research,  in optics, acoustics, plasma, hydrodynamics, particle physics, and others.
Depending on the nature of the nonlinearity, various mechanisms of the formation of nonlinear solitary waves are realized. Nevertheless, the general properties of solitary waves in different materials and physical situations can be similar. Usually, two main types of solitary waves are considered: single-component (scalar) and two-component (vector) nonlinear waves. Among single-component solitary waves, soliton and breather are very often investigated. But the  breather is of special interest because can be propagated also for relatively low intensity of the wave  in comparison to soliton. They have been studied for a very long time both theoretically and experimentally in a lot of nonlinear physical phenomena \cite{Newell::85, Allen::75, Sauter::96, Adamashvili:Phys.Rev.A:19, Crisp:PhysRep:70, Rothenberg::1984, Arkhipov::2016,Adamashvili:Phys.Rev.A:17, Diels:PhysRev:74,Harvey::94, Arkhipov::2018, Adamashvili:OptLett:06,Adamashvili:PhysRevE:06}. In addition to single-component scalar breathers, two-component waves, such as vector solitons, are also considered in various physical systems, for example, in optical waveguides \cite{Menyuk::2010, KivAg::03}. Recently, in optics, and then in acoustics, some special varieties of two-component waves, which are the bound state of two small amplitude scalar one-component breathers with the same polarization, have been investigated. One breather oscillates with the sum, and the second with difference of the frequencies and wavenumbers (SDFW). This nonlinear two-component wave has a very special profile which differs in shape of all other nonlinear solitary waves (see, Fig.1). In the theory of self-induced transparency, such nonlinear solitary wave is called the vector  $0\pi$ pulse \cite{Adamashvili:Result:11,Adamashvili:OS:19,Adamashvili:PhysRevE:12, Adamashvili:Optics and spectroscopy:2012, Adamshvili:Arxiv:2019}.

The single-component and two-component solitary waves behavior can mathematically be described by means of the nonlinear partial  differential equations. Among them are the Sin-Gordon equation (SGE), the Maxwell-Bloch equations, the Maxwell-Liouville equations, the wave equation in Kerr media, the system of  wave equation and material equations for two-photon resonant transitions, the system of the magnetic Bloch equations and the elastic wave equation, and many others.
The inverse scattering transform (IST), the perturbative reduction method (PRM) and a lot of other approaches, in solving the nonlinear differential equations  and to  analyze the solitary waves have been  applied \cite{Newell::85,Novikov::84,Dodd::1982,Ablowitz::81,Ablowitz1::73, Leblond::08,Taniuti::1973}.
In particular, a small amplitude breather solution of SGE can be obtained by means of IST (see, Appendix).
The two-component vector breathers oscillating with SDFW have been considered for optical and acoustic waves by means of the Maxwell-Bloch equations, the Maxwell-Liouville equations, SGE and others,  using recently developed the generalized version of the PRM \cite{Adamashvili:Result:11, Adamashvili:Optics and spectroscopy:2012, Adamashvili:OS:19,Adamashvili:PhysRevE:12,Adamashvili:Physica B:14,Adamashvili:PhysLettA:2015}.

Besides with the above mentioned areas of physics and corresponding nonlinear differential equations, the various nonlinear solitary waves  in absolutely other  physical systems can be described by means of the nonlinear dispersive  modified Benjamin-Bona-Mahony (MBBM) equation.   It can describes an approximations for surface long waves in  dispersive nonlinear materials, the  phonons properties in anharmonic crystal lattice, acoustic-gravity waves in compressible fluids, hydro-magnetic waves in cold plasma \cite{Benjamin::1972, Guner::17, Manafianheris::12, Riskin ::2010}.

The nonlinear dispersive MBBM equation is given by
\begin{equation}\label{mbbm}
u_{ t}+C u_{z}+\beta u_{zzz} +a u^{2} u_{z}=0,
\end{equation}
or in the dimensionless form \cite{Khorshidi::16}
\begin{equation}\label{bbm1}\nonumber\\
u_{ t}+ u_{z}+ u_{zzz} +a u^{2} u_{z}=0,
\end{equation}
where $u(z,t)$ is a real function of space and time, $z$ is spatial variable and $t$ is time variable. $C$, $\beta$ and $a$ are arbitrary constants.

Eq.\eqref{mbbm} have been considered as an improvement of the modified Korteweg-de Vries equation which is also well known in different field of research and applications \cite{Guner::17, Riskin ::2010, Khorshidi::16, Triki::12,  Wazwaz::17}.

Sometimes another form of the MBBM equation is also considered \cite{Ghanbari::19, Khater::17}
\begin{equation}\label{bbm}\nonumber\\
u_{ t}+C u_{z}+\beta u_{zzt} +a u^{2} u_{z}=0.
\end{equation}

The linear part of MBBM equation \eqref{mbbm} is given by
\begin{equation}\label{bbml}\nonumber\\
u_{ t}+C u_{z}+\beta u_{zzz} =0,
\end{equation}
which describes dispersive effect to the equation and yields dispersion relation  between the wavenumber $k$ and frequency $\omega$:
\begin{equation}\label{dis14}
\omega=C k- \beta   k^{3}.
\end{equation}

In the system of coordinates moving along the axis $z$ with velocity $C$, the equation \eqref{mbbm} is transformed to the modified Kortrweg-de-Vries equation
\begin{equation}\label{kdv}
u_{ t}+\beta u_{yyy} +a u^{2} u_{y}=0,
\end{equation}
where
\begin{equation}\label{y}
y=z-Ct,\;\;\;\;\;\;\;\;\;\;\;\;t=t.
\end{equation}
Eq.\eqref{kdv} has been proposed as model to describe the nonlinear evolution of plasma waves.

Although the properties of various solitary waves of the MBBM equation have been studied, a two-component breather with the SDFW has not  been considered up to now. The purpose of the present work is to theoretically investigate the  two-component breather solution with the SDFW of the MBBM equation \eqref{mbbm} by using  the generalized PRM and comparison with the single-component breather of this equation analyzed by  the PRM.

The rest of this paper is organized as follows. Section II is devoted to the  two-component (vector) breather solution with the SDFW of the MBBM equation by using the generalized version of the PRM.  We will investigate the one-component (scalar) breather solution of the MBBM equation  by the PRM in standard form in the Section III.  Finally, in the last Section IV, we will comparison two methods: PRM in standard form and the generalized PRM and corresponding solutions, and also discuss obtained results.

\section{Vector breather and the generalized PRM}

In the beginning we will simplify Eq.\eqref{mbbm} by using the method of slowly varying envelope approximation in the following form \cite{Riskin ::2010, Vinogradova::90}:
\begin{equation}\label{uz}
 u=\sum_{l=\pm1}\hat{u}_{l}Z_{l},
\end{equation}
where $Z_{l}= e^{il(kz -\om t)}$ is the fast oscillating part of the carrier wave and $\hat{u}_{l}$ are the slowly varying complex
amplitudes which satisfied inequalities
\begin{equation}\label{swa}
 \left|\frac{\partial \hat{u}_{l}}{\partial t}\right|\ll\omega
|\hat{u}_{l}|,\;\;\;\left|\frac{\partial \hat{u}_{l}}{\partial z
}\right|\ll k|\hat{u}_{l}|.
\end{equation}
To guarantee the reality of the function $u$, we set $\hat{u}_{l}=\hat{u}_{-l}^{*}$.

Substituting Eq.\eqref{uz} into \eqref{mbbm} we obtain the equation for the envelope functions $\hat{u}_{l}$:
\begin{equation}\label{en}
 \sum_{l=\pm1}[\frac{\pa \hat{u}_{l}}{\pa t}
 +( C -3\beta k^2 l^2 )\frac{\pa \hat{u}_{l}}{\pa z}
 + 3\beta ilk \frac{\pa^{2} \hat{u}_{l}}{\pa z^2}
 +\beta \frac{\pa^{3} \hat{u}_{l}}{\pa z^3}
  ]  Z_{l}=-a \sum_{L,m,l'} Z_{L+m+l'}       \hat{u}_{L} \hat{u}_{m}(il'k  \hat{u}_{l'}+\frac{\pa \hat{u}_{l'}}{\pa z}).
\end{equation}
and the connection between parameters $\omega$ and $k$ described by Eq. \eqref{dis14}.

In order to study the two -component nonlinear wave solution of the Eq.\eqref{mbbm} we use the generalized PRM developed in Refs.\cite{Adamashvili:Result:11,Adamashvili:Eur.Phys.J.D.:12,Adamashvili:Optics and spectroscopy:2012,Adamshvili:Arxiv:2019} which makes it possible to transform the MBBM equation for slowly envelope functions $\hat{u}_{l}$ Eq.\eqref{en} to the coupled NSEs for auxiliary functions $f_{l,n}^ {(\alpha)}$.
As a result, we will obtain two-component vector breather with the SDFW. Using this method, the complex envelope function  $\hat{u}_{l}$ can be represented in the form
\begin{equation}\label{gprm}
\hat{u}_{l}(z,t)=\sum_{\alpha=1}^{\infty}\sum_{n=-\infty}^{+\infty}\varepsilon^\alpha
Y_{l,n} f_{l,n}^ {(\alpha)}(\zeta_{l,n},\tau),
\end{equation}
where $\varepsilon$ is a small parameter,
$$
Y_{l,n}=e^{in(Q_{l,n}z-\Omega_{l,n}
t)},\;\;\;\zeta_{l,n}=\varepsilon Q_{l,n}(z-{v_{g;}}_{l,n} t),\;\;\;\tau=\varepsilon^2 t,\;\;\;
{v_{g;}}_{l,n}=\frac{d\Omega_{l,n}}{dQ_{l,n}}.
$$

Substituting Eq.\eqref{gprm} into Eq.\eqref{en} we obtain
\begin{equation}\label{rr51}
 \sum_{l=\pm 1}\sum_{\alpha=1}^{\infty}\sum_{n=-\infty}^{+\infty}\varepsilon^\alpha Z_{l} Y_{l,n} [
\mathfrak{W}_{l,n}f_{l,n}^{(\alpha)} +\varepsilon \mathfrak{J}_{l,n}  \frac{\partial f_{l,n}^{(\alpha)}}{\partial
\zeta_{l,n}}  +3\beta i \varepsilon^{2} Q^{2}_{l,n} ( l k    +    n Q_{l,n} )  \frac{\partial^{2} f_{l,n}^{(\alpha)}}{\partial
\zeta^{2}_{l,n}}+\varepsilon^2 \frac{\partial f_{l,n}^{(\alpha)}}{\partial \tau} +  O(\varepsilon^3)] +$$$$ \varepsilon^3 i a \sum_{L,m,l',N,N'=\pm1}     Z_{L+m+l'}  Y_{L,N} Y_{m,N'}  f_{L,N}^ {(1)}
  f_{m,N'}^ {(1)}  [(l'k + Q_{l',+1}) Y_{l',+1} f_{l',+1}^{(1)}
  +(l'k - Q_{l',-1})Y_{l',-1} f_{l',-1}^{(1)}]+  O(\varepsilon^4)=0,
 \end{equation}
 where
\begin{equation}\label{eqoo}
\mathfrak{W}_{l,n}  =- in( \Omega_{l,n}- C Q_{l,n} +3\beta k^2 l^2  Q_{l,n} + 3 \beta l n k  Q_{l,n}^{2} + \beta n^2 Q_{l,n}^{3}),
$$$$
\mathfrak{J}_{l,n}= Q (- {v_{g;}}_{l,n} +C -3\beta k^2 l^2 - 6\beta lk  n Q_{l,n}- 3 \beta n^2 Q_{l,n}^{2} ).
\end{equation}

There are four independent equations for different values $l=\pm1$ and $n=\pm1$.

Equating to zero, the terms with the same powers of a small parameter  $\varepsilon$, we obtain a series of equations. In the first order of a small parameter $\varepsilon$, we obtain the following equation
\begin{equation}\label{eqvb5}
 \sum_{l=\pm 1}\sum_{n=-\infty}^{+\infty} Z_{l} Y_{l,n} \mathfrak{W}_{l,n}f_{l,n}^{(1)}=0.
\end{equation}

From Eq.\eqref{eqvb5} we can established connection between different values of the parameters $\Omega_{l,n}$ and $Q_{l,n}$. In particular, when
\begin{equation}\label{dis1}
 \Omega_{\pm1,\pm1}-C Q_{\pm1,\pm1} +3\beta k^2 Q_{\pm1,\pm1}  + 3\beta  k  Q^{2}_{\pm1,\pm1}+ \beta Q^{3}_{\pm1,\pm1}=0,
 \end{equation}
then $f_{\pm1,\pm1}^{(1)}\neq0$. But in case when
\begin{equation}\label{dis2}
\Omega_{\pm1,\mp1}-C Q_{\pm1,\mp1} +3\beta k^2 Q_{\pm1,\mp1}  - 3\beta  k  Q^{2}_{\pm1,\mp1}+ \beta Q^{3}_{\pm1,\mp1}=0,
\end{equation}
then $f_{\pm1,\mp1}^{(1)}\neq0.$

From Eq.\eqref{rr51}, in the second order of the $\varepsilon$, we obtain the following equation
\begin{equation}\label{eqoo}
 \sum_{l=\pm 1}\sum_{n=-\infty}^{+\infty} Z_{l} Y_{l,n}[ \mathfrak{W}_{l,n} f_{l,n}^{(2)}+ \mathfrak{J}_{l,n}\frac{\partial f_{l,n}^{(1)}}{\partial \zeta_{l,n}}]=0.
\end{equation}
To take into account Eqs.\eqref{dis1} and \eqref{dis2}, from the Eq.\eqref{eqoo}  follows
$$
\mathfrak{J}_{\pm1,\pm1}=\mathfrak{J}_{\pm1,\mp1}=0,\;\;\;\;\;\;\;\;
 f_{+1,\pm2}^{(2)}=f_{-1,\pm2}^{(2)}=0,
$$
and
\begin{equation}\label{dis11}
 {v_{g;}}_{\pm1,\pm1}=\frac{d \Omega_{\pm1,\pm1}}{dQ_{\pm1,\pm1}}=C  -3\beta k^2  -6 \beta  k  Q_{\pm1,\pm1}-3 \beta Q^{2}_{\pm1,\pm1},
$$$$
 {v_{g;}}_{\pm1,\mp1}=\frac{d \Omega_{\pm1,\mp1}}{dQ_{\pm1,\mp1}}= C -3\beta k^2  + 6\beta  k  Q_{\pm1,\mp1}- 3\beta Q^{2}_{\pm1,\mp1}.
\end{equation}

From the equation \eqref{rr51} we have in the third order of the $\varepsilon$ the system of equations proportional $Z_{+1}$ and $Z_{-1}$, respectively
\begin{equation}\label{pp}
-i \frac{\partial  f_{+1,\pm1}^{(1)}}{\partial \tau}+ 3\beta  Q^{2}_{+1,\pm1} (k \pm Q_{+1,\pm1}) \frac{\partial^{2}  f_{+1,\pm1}^{(1)}}{\partial \zeta^{2}_{+1,\pm1}} +  a   (k \pm Q_{+1,\pm1}) [  |f_{+1,\pm1}^{(1)}|^{2} + 2 |f_{+1,\mp1}^ {(1)}|^{2} ]f_{+1,\pm1}^{(1)}=0,
\end{equation}
\begin{equation}\label{mm}\nonumber\\
 i \frac{\partial  f_{-1,\mp1}^{(1)}}{\partial \tau}+   3  \beta  Q^{2}_{-1,\mp1} (k\pm Q_{-1,\mp1}) \frac{\partial^{2}  f_{-1,\mp1}^{(1)}}{\partial \zeta^{2}_{-1,\mp1}}  + a  (k  \pm Q_{-1,\mp1})  (  |f_{-1,\mp1}^ {(1)}|^{2}   +2   |f_{-1,\pm1}^ {(1)}|^{2}  ) f_{-1,\mp1}^{(1)}=0.
\end{equation}
We consider the equations proportional to the $Z_{+1}$ in detailed, the complex-conjugation equations proportional to the $Z_{-1}$ can be considered similarly.

After transformation back to the variables $z$ and $t$, from the Eq.\eqref{pp} we obtain
\begin{equation}\label{pp2}
i (\frac{\partial U_{+1}}{\partial t}+v_{+}\frac{\partial  U_{+1}} {\partial z})+p_{+} \frac{\partial^{2} U_{+1} }{\partial z^{2}} +q_{+}  |U_{+1}|^{2}U_{+1} + r_{+} |U_{-1}|^{2} U_{+1}=0,
$$
$$
i (\frac{\partial U_{-1}}{\partial t}+ v_{-} \frac{\partial U_{-1}}{\partial z})+p_{-} \frac{\partial^{2} U_{-1} }{\partial z^{2}}
   +q_{-}  |U_{-1}|^{2} U_{-1} + r_{-} |U_{+1}|^{2}U_{-1} =0,
\end{equation}
where
\begin{equation}\label{vpqr}
U_{\pm1}=\varepsilon  f_{+1,\pm1}^{(1)},
$$$$
v_{\pm}=v_{{g;}_{+1,\pm1}},\;\;\;\;\;\;\;\;\;\;\;\;\;p_{\pm}=-3\beta  (k \pm Q_{+1,\pm1}),
$$$$
q_{\pm}= -  a   (k  \pm Q_{+1,\pm1}),\;\;\;\;\;\;\;\;\;\;\;\;\;\;\;r_{\pm}=2q_{\pm},
$$$$
\Omega_{+1,+1}=\Omega_{-1,-1}=\Omega_{+1},\;\;\;\;\;\;\;\;\;\;\;\Omega_{+1,-1}=\Omega_{-1,+1}=\Omega_{-1},
$$$$
Q_{+1,+1}=Q_{-1,-1}=Q_{+1},\;\;\;\;\;\;\;\;\;\;\;Q_{+1,-1}=Q_{-1,+1}=Q_{-1}.
\end{equation}

Eqs.\eqref{pp2} is well known the coupled NSEs which are characterizing bound state of the two breathers. One breather is described by means of the function $U_{+1}$
and the second breather by the function $U_{-1}$. These breathers are propagating in the same direction, with the same constant velocities and  have identical polarizations. In the process of propagation they interact to each other and exchange energy. This connection between breathers is characterized by the cross terms in Eqs.\eqref{pp2},  $r_{+} |U_{-1}|^{2} U_{+1}$  and $r_{-} |U_{+1}|^{2}U_{-1}$. As result, we obtain one steady-state two-component vector pulse which consists from the two breathers. In order to describe properties of the vector two-component pulse we have to find solution of the coupled NSEs for the auxiliary functions $U_{+1}$ and  $U_{-1}$.

We will seek the steady-state solutions of Eqs.\eqref{pp2} for functions $U_{\pm1}$ in the following form
\begin{equation}\label{uu2}
U_{\pm}=K_{\pm}\; S( \xi )e^{i\varphi_{\pm}},
\end{equation}
where $\varphi_{\pm}=k_{\pm} z- \omega_{\pm} t$ are the phase functions for each components, $K_{\pm},\;k_{\pm}$ and $\omega_{\pm}$ are constants.
The standard way to ensure the stationary character of the envelope function $S( \xi )$ is to require to depend on space coordinate and time only through the variable $ \xi = t- \frac{z}{V_{0}}$, where $V_{0}$ is the nonlinear two-component pulse velocity. We suppose that valid following inequalities
\begin{equation}\label{sko}
k_{\pm}<<Q_{\pm 1},\;\;\;\;\omega_{\pm}<<{\Omega}_{\pm 1}.
\end{equation}

Substituting Eqs.\eqref{uu2}  into Eqs.\eqref{pp2}  we obtain the explicit form of the envelope function and connections between different parameters of the breathers
\begin{equation}\label{rt16}
S(\xi)=\frac{1}{\mathfrak{b} T}sech(\frac{t- \frac{z}{V_{0}}}{T}),
$$$$
K_{+}^{2}=\frac{p_{+}q_{-}-2 p_{-} q_{+}}{p_{-}q_{+}-2p_{+} q_{-}}K_{-}^{2},
$$$$
\omega_{+}=\frac{p_{+}}{p_{-}}\omega_{-}+\frac{V^{2}_{0}(p_{-}^{2}-p_{+}^{2})+v_{-}^{2}p_{+}^{2}-v_{+}^{2}p_{-}^{2}
}{4p_{+}p_{-}^{2}},\;\;\;\;\;k_{\pm}=\frac{V_{0}-v_{\pm}}{2p_{\pm}}.
\end{equation}

Substituting Eqs.\eqref{uu2} and \eqref{rt16} into the Eqs.\eqref{gprm} and \eqref{uz} we obtain the two-component vector breather with the SDFW of the MBBM equation in the form
\begin{equation}\label{myi}
u(z,t)=
\frac{2}{\mathfrak{b} T}sech(\frac{t-\frac{z}{V_{0}}}{T})\{  K_{+} \cos[(k+Q_{+1}+k_{+})z -(\om +\Omega_{+1}+\omega_{+}) t]+
$$$$
K_{-}\cos[(k-Q_{-1}+k_{-})z -(\om -\Omega_{-1}+\omega_{-})t]\},
\end{equation}
where T is the width of the two-component nonlinear pulse,
\begin{equation}\label{er17}
T^{-2}=V_{0}^{2}\frac{v_{+}k_{+}+k_{+}^{2}p_{+}-\omega_{+}}{p_{+}},
$$$$
\mathfrak{b}^{2}=\frac{V_{0}^{2} q_{+}}{2p_{+}}(K_{+}^{2}+2 K_{-}^{2}) .
\end{equation}

Eq.\eqref{myi} is described the nonlinear two-component vector pulse (vector breather) which consists from the two breathers. One of them is oscillating with the sum  of the frequencies $\om +\Omega_{+1}$ and wavenumbers $k+Q_{+1} $(to take into account Eqs. \eqref{sko}) and the second breather is oscillating with the difference of the frequencies $\om -\Omega_{-1}$ and wavenumbers $k-Q_{-1} $.  The parameters of the nonlinear wave by the equations \eqref{dis11}, \eqref{vpqr}, \eqref{rt16}, and \eqref{er17} are determined. Corresponding plot of the two-component vector breather with the SDFW of the MBBM equation  at a fixed value of the z coordinate for the values of the parameters $\omega/ \Omega_{\pm 1}=10^{3},\;\; \Omega_{+1}/\Omega_{-1}=0.83$ in the Fig.1 is presented.

\begin{figure}[htbp]
\includegraphics[width=0.44\textwidth]{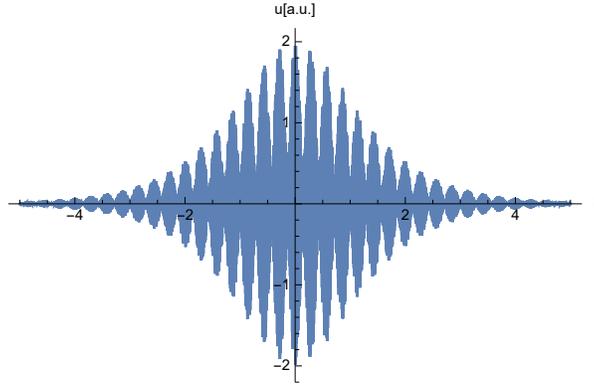}
\caption{ Plot of the two-component breather with the SDFW  of the MBBM equation \eqref{myi}
 at a fixed value of the z coordinate.}
\label{fig1}
\end{figure}

\section{One-component breather }

In the present section, we will consider the one-component breather solution of the MBBM equation \eqref{mbbm} by the PRM.  The function $u$ is expanded in a power series of $\varepsilon$ as \cite{Taniuti::1973}
\begin{equation}\label{prm}
u(z,t)=\sum_{\alpha=1}^{\infty}\sum_{n=-\infty}^{+\infty}\varepsilon^\alpha
Y_{n} \mathfrak{f}_{n}^ {(\alpha)}(\zeta,\tau),
\end{equation}
where $\mathfrak{f}_{n}^ {(\alpha)}(\zeta,\tau)$ is the slowly varying complex envelope function,
$$
Y_n=e^{in(\kappa z-w t)}
$$
is the rapidly varying part, while the slow variables
$$
\zeta=\varepsilon \kappa(z-v_g
t),\;\;\;\tau=\varepsilon^2 t,\;\;\;\  v_g=\frac{dw}{d\kappa},
$$

We  assume that they satisfy the inequalities
$$
\left|\frac{\partial
\mathfrak{f}_{n}^{(\alpha )}}{
\partial t}\right|\ll w \left|\mathfrak{f}_{n}^{(\alpha)}\right|,\;\;\left|\frac{\partial
\mathfrak{f}_{n}^{(\alpha )}}{\partial \eta }\right|\ll \kappa \left|\mathfrak{f}_{n}^{(\alpha )}\right|.
$$
We use the same notation as before for small parameter $\varepsilon$. To guarantee the reality of the function $u$, we set $\mathfrak{f}_{n}^{(\alpha)}=\mathfrak{f}_{-n}^{(\alpha){*}} $.

Substituting the expansion  \eqref{prm}  into  Eq.\eqref{mbbm}, we obtain  the equations
\begin{equation}\label{wlh}
\sum_{\alpha=1}^{\infty}\sum_{n=-\infty}^{+\infty}\varepsilon^\alpha
Y_n [-in(w-C\kappa+ n^2 \beta \kappa^{3}) -\varepsilon \kappa (v_g -C+ 3\beta n^2 \kappa^{2})  \frac{\partial }{\partial \zeta} + \varepsilon^{2} 3\beta i n \kappa^{3}   \frac{\partial^{2} }{\partial \zeta^{2}} +\varepsilon^2
\frac{\partial }{\partial \tau}  +  O(\varepsilon^4)) \mathfrak{f}_{n}^{(\alpha)}$$$$+ \varepsilon^{3} a i  \kappa   \sum_{m,m_{1},m_{2}=-\infty}^{+\infty}
 Y_{m+m_{1}+m_{2}} m_{2}  \mathfrak{f}_{m}^ {(1)} \mathfrak{f}_{m_{1}}^{(1)}\mathfrak{f}_{m_{2}}^{(1)}=0.
\end{equation}

To determine the functions of $\mathfrak{ f}_{n}^{(\alpha)}$, we equate to zero the different terms corresponding to same powers of $\varepsilon$.

In first order in $\varepsilon$, we obtain
\begin{equation}\label{py}
\sum_{n=-\infty}^{+\infty}\varepsilon Y_n [-in(w-C\kappa+ n^2 \beta \kappa^{3}) ] \mathfrak{f}_{n}^{(1)}=0.
\end{equation}
From Eq.\eqref{py} we obtain the dispersion relation
\begin{equation}\label{dd2}
w=C\kappa - \beta \kappa^{3}.
\end{equation}
The following components of the function $\mathfrak{ f}_{n}^{(1)}$ can differ from zero: $\mathfrak{ f}_{+1}^{(1)}$ and $ \mathfrak{f}_{-1}^{(1)}$, and the same time $\mathfrak{f}_{n\neq\pm1}^{(1)}=0$.

To second order in  $\varepsilon$, we obtain the equation
\begin{equation}\label{vg}
v_g=C- 3 \beta \kappa^{2}.
\end{equation}

To take into account Eqs. \eqref{dd2} and \eqref{vg}, to third order in $\varepsilon$, from the Eq.\eqref{wlh} we finally obtain the equations for the functions  $\mathfrak{ f}_{\pm1}^{(1)}$
\begin{equation}\label{ns}
 \mp i \frac{\partial  \mathfrak{f}_{\pm 1}^{(1)}}{\partial \tau}+   3 \beta  \kappa^{3}   \frac{\partial^{2}  \mathfrak{f}_{\pm 1}^{(1)}}{\partial \zeta^{2}}  +a \kappa  |\mathfrak{f}_{\pm 1}^{(1)}|^{2} \mathfrak{f}_{\pm 1}^{(1)}=0
\end{equation}

Using the expressions
$$
\frac{\partial }{\partial \zeta}=\frac{1}{\varepsilon \kappa}\frac{\partial
}{\partial z}, \;\;\;\;\;\;\;\;\;\;\;\;\;\;\;\frac{\partial
}{\partial \tau}=\frac{v_{g}}{\varepsilon^2}\frac{\partial }{\partial
z}+\frac{1}{\varepsilon^{2}}\frac{\partial }{\partial t},
$$
we can transform the equations \eqref{ns} to the form
\begin{equation}\label{nse}
-l i( v_{g} \frac{\partial \lambda_{l}}{\partial
z}+\frac{\partial \lambda_{l}}{\partial t}) + 3\beta  \kappa \frac{\partial^2 \lambda_{l}
}{\partial z^2}   +a \kappa |\lambda_{l}|^{2} \lambda_{l}=0
\end{equation}
where
 $$
 \lambda_{l}=\varepsilon \mathfrak{f}_{l}^{(1)},\;\;\;\;\;\;\;\;\;\;\;\;l=\pm1.
 $$

If we make the transformation of the variables Eq.\eqref{y}, from the Eq.\eqref{nse} we finally obtain the NSE in the form
\begin{equation}\label{nse1}
-l i\frac{\partial \lambda_{l}}{\partial t} + p \frac{\partial^2 \lambda_{l}
}{\partial y^2}   +q |\lambda_{l}|^{2} \lambda_{l}=0
\end{equation}
where
\begin{equation}\label{pq}
p=3 \beta \kappa,\;\;\;\;\;\;\;\;\;\;\;q=a \kappa.
\end{equation}

Eq.\eqref{nse1} has the soliton solution \cite{Novikov::84, Dodd::1982, Ablowitz::81, Ablowitz1::73}
\begin{equation}\label{soli}
\lambda_{l}= K\frac{e^{- i l \phi_1}} {cosh\phi_2},
\end{equation}
where
\begin{equation}\label{phi}
\phi_1= \frac{V_b}{6 \beta \kappa}z -[\frac{V_{b}
}{6 \beta \kappa}(v_{g}+\frac{V_{b}}{2})-\frac{a \kappa}{2}K^{2}]t,
$$$$
\phi_2={K\sqrt{\frac{a}{6 \beta}}[z- (v_{g}+V_{b})t]}.
\end{equation}
$K$  is the amplitude of the NSE soliton, $V_b$ is the velocity of the nonlinear wave, $\lambda_{l}$ is the envelope of the soliton.

Substituting the soliton solution for  $\lambda_{l}$, Eq.\eqref{soli}, into Eq.\eqref{prm}, we obtain for the envelope for $u$ the one-component (scalar) breather solution of the MBBM equation in the form
\begin{equation}\label{br}
u(z,t)=2 K\frac{sin { (\kappa z-w t-\phi _{1}+\frac{\pi}{2})}}{\cosh \phi_{2}} +O(\varepsilon^2).
\end{equation}

From \eqref{br}, it is obvious that the trigonometric function points of the existance of oscillation and this leads to the fact that the soliton solution of the NSE \eqref{soli} for the auxiliary function $\lambda_{l}$ is transformed to the one-component breather solution \eqref{br} of the MBBM equation for the function $u(z,t)$.
The profile and parameters of the one-component breather of MBBM equation are determined from the Eqs.\eqref{pq},\eqref{phi} and \eqref{br}.The connection between parameters
$w$ and $\kappa$ is determined from the Eq.\eqref{dd2}.

From the Eqs. \eqref{br} and \eqref{u2} obvious that the one-component breather solution of MBBM equation is coincide with the small amplitude breather of SGE with precision to the notation (see, Appendix).

\section{Conclusion }

We consider the formation of nonlinear waves in various physical systems (anharmonic crystals, hydrodynamics, plasma, etc.) which are described by the MBBM equation \eqref{mbbm}.

Using a recently developed mathematical approach, in particular, the generalized PRM, it became possible to obtain a new solution of the MBBM equation, which is a nonlinear combination of the components of two breathers. One component oscillates with the sum, and the second with the difference of the frequencies and wavenumbers. Both components have the same polarization, and as a result of superposition of their amplitudes, a nonlinear pulse with a specific profile is obtained. Although the analytical expression of the pulse Eq.\eqref{myi} differs from the analytical expression of the $0\pi$ pulse of the self-induced transparency studied at the beginning in nonlinear optics, and later in nonlinear acoustics, the corresponding graphical expressions are similar (see, Fig. 1)\cite{Adamashvili:Result:11, Adamashvili:Optics and spectroscopy:2012, Adamashvili:OS:19, Adamashvili:PhysRevE:12}.

We also obtained a one-component breather solution of the MBBM equation by means of PMR in a standard form. This wave is completely coincides with the  low-amplitude breather solution of  SGE. Comparison of analytical expressions and corresponding profiles of the two-component and one-component breathers of the MBBM equation, it becomes obvious that they are completely different.

The study of the two-component breather solutions  became possible due to the generalized PRM, which allows us to expand the number of auxiliary functions and parameters compared to standard PRM. In particular, in the generalized PRM we use two complex functions  $f_{\pm1,\pm1}$ and $f_{\pm1, \mp1}$ and eight parameters $\Omega_{l,n}$ and $Q_{l,n}$, when in the standard PMR there is only one complex function $\mathfrak{f}_{n}$ and two parameters   $w$ and $\kappa$ which is not enough to describe two-component waves.

The results obtained above allow us to make an important conclusion that the generalized PRM allows us to get a new solution Eq.\eqref{myi} (Fig. 1) of the MBBM equation \eqref{mbbm}. Such a nonlinear two-component pulse previously studied in optics and acoustics, can also be formed in other physical systems, in particular, in anharmonic crystals, hydrodynamics, plasma, etc. Present  work significantly extends the class of materials and physical situations in which vector breathers with SDFW can be formed. Consequently, existence of the vector breathers with SDFW has a general character and can be met in completely different physical systems and various physical situations.

\vskip+0.5cm
\centerline{\textbf{Appendix }}

The standard definition of breather  is a  two-soliton solution of the SGE which is stable solution to relatively infinitesimal perturbations of the initial data. The breather solution of SGE can be obtained by the inverse scattering transform and written as \cite{Novikov::84, Dodd::1982, Ablowitz::81, Ablowitz1::73}
\begin{equation}\label{b}
\arctan {A\frac{sin \varphi_{1}}{cosh \varphi_{2}}},
\end{equation}
where A is the amplitude of the wave, the functions $\varphi_{1}$ and $\varphi_{1}$ are determined  in Ref.\cite{ Novikov::84}. When $A<<1$, the breather in Eq.\eqref{b}  can be reduced to a small amplitude breather
\begin{equation}\label{u2}
A \frac{sin \varphi_{2}}{cosh \varphi_{2}}.
\end{equation}

\vskip+0.5cm


\begin{thebibliography}{20}
\expandafter\ifx\csname natexlab\endcsname\relax\def\natexlab#1{#1}\fi
\expandafter\ifx\csname bibnamefont\endcsname\relax

\def\bibnamefont#1{#1}\fi
\expandafter\ifx\csname bibfnamefont\endcsname\relax
\def\bibfnamefont#1{#1}\fi
\expandafter\ifx\csname citenamefont\endcsname\relax
\def\citenamefont#1{#1}\fi
\expandafter\ifx\csname url\endcsname\relax
\def\url#1{\texttt{#1}}\fi
\expandafter\ifx\csname urlprefix\endcsname\relax\def\urlprefix{URL }\fi
\providecommand{\bibinfo}[2]{#2}
\providecommand{\eprint}[2][]{\url{#2}}

\bibitem[{\citenamefont{Newell}(1975)}]{Newell::85}
\bibinfo{author}{\bibfnamefont{A.~C.}~\bibnamefont{Newell}},
\emph{\bibinfo{title}{\emph{Solitons in Mathematics and Physics }}}
(\bibinfo{publisher}{Society for Industrial and Applied Mathematics}, \bibinfo{year}{1985}).

\bibitem[{\citenamefont{Allen and Eberly}(1975)}]{Allen::75}
\bibinfo{author}{\bibfnamefont{L.}~\bibnamefont{Allen}} \bibnamefont{and}
\bibinfo{author}{\bibfnamefont{J.}~\bibnamefont{Eberly}},
\emph{\bibinfo{title}{Optical resonance and two level atoms}}
(\bibinfo{publisher}{Dover}, \bibinfo{year}{1975}).

\bibitem[{\citenamefont{Sauter E.G.}(1996}]{Sauter::96}
\bibinfo{author}{\bibfnamefont{E.~G.} \bibnamefont{Sauter}},
\emph{\bibinfo{title}{Nonlinear Optics}}
(\bibinfo{publisher}{Wiley, New York,} \bibinfo{year}{1996}).

\bibitem[{\citenamefont{Crisp}(1990)\citenamefont{Crisp}}]{Crisp:PhysRep:70}
\bibinfo{author}{\bibfnamefont{M.~D.} \bibnamefont{Crisp}},
\bibinfo{journal}{Phys. Rev. A.}
  \textbf{\bibinfo{volume}{2}}, \bibinfo{pages}{2172} (\bibinfo{year}{1970}).

\bibitem[{\citenamefont{Rothenberg}(1973)}]{Rothenberg::1984}
\bibinfo{author}{\bibfnamefont{J.~E.} \bibnamefont{Rothenberg}},
\bibinfo{author}{\bibfnamefont{D.}~\bibnamefont{Grischkowsky}} \bibnamefont{and}
\bibinfo{author}{\bibfnamefont{A.~C.}\bibnamefont{Balant}},
\bibinfo{journal}{Phys. Rev. Lett.} \textbf{\bibinfo{volume}{53}},
\bibinfo{pages}{552} (\bibinfo{year}{1984}).

\bibitem[{\citenamefont{Adamashvili and Kaup}(2006)}]{Adamashvili:PhysRevE:06}
\bibinfo{author}{\bibfnamefont{G.~T.} \bibnamefont{Adamashvili}}
  \bibnamefont{and} \bibinfo{author}{\bibfnamefont{D.~J.}~\bibnamefont{Kaup}},
  \bibinfo{journal}{Phys. Rev. E},  \textbf{\bibinfo{volume}{73}},
  \bibinfo{pages}{066613} (\bibinfo{year}{2006}).

\bibitem[{\citenamefont{Arkhipov}(1973)}]{Arkhipov::2016}
\bibinfo{author}{\bibfnamefont{R.~M.} \bibnamefont{Arkhipov}},
\bibinfo{author}{\bibfnamefont{M ~V.} \bibnamefont{Arkhipov}},
\bibinfo{author}{\bibfnamefont{I. }~\bibnamefont{Babushkin}} \bibnamefont{and}
\bibinfo{author}{\bibfnamefont{N.~N.} \bibnamefont{Rosanov }},
\bibinfo{journal}{Optics lett.} \textbf{\bibinfo{volume}{41}},
\bibinfo{pages}{737} (\bibinfo{year}{2016}).

\bibitem[{\citenamefont{ Harvey}(1994)}]{Harvey::94}
\bibinfo{author}{\bibfnamefont{J.~D.} \bibnamefont{ Harvey}},
\bibinfo{author}{\bibfnamefont{J.~M.} \bibnamefont{Dudley}},
\bibinfo{author}{\bibfnamefont{P.~F.} \bibnamefont{Curley}},
\bibinfo{author}{\bibfnamefont{C.}~\bibnamefont{Spielmann}} \bibnamefont{and}
\bibinfo{author}{\bibfnamefont{F.}~\bibnamefont{Krausz  }},
\bibinfo{journal}{Optics lett.} \textbf{\bibinfo{volume}{19}},
\bibinfo{pages}{972} (\bibinfo{year}{1994}).

\bibitem[{\citenamefont{Adamashvili }(2017)}]{Adamashvili:Phys.Rev.A:17}
\bibinfo{author}{\bibfnamefont{G.~T.} \bibnamefont{Adamashvili}} \bibnamefont{and}
\bibinfo{author}{\bibfnamefont{D.~J.} \bibnamefont{Kaup}},
\bibinfo{journal}{Phys. Phys. A.}  \textbf{\bibinfo{volume}{95}},
\bibinfo{pages}{053801} (\bibinfo{year}{2017}).

\bibitem[{\citenamefont{Arkhipov}(2018)}]{Arkhipov::2018}
\bibinfo{author}{\bibfnamefont{M.~V.} \bibnamefont{Arkhipov}},
\bibinfo{author}{\bibfnamefont{A.~A.} \bibnamefont{Shimko}},
\bibinfo{author}{\bibfnamefont{R.~M.} \bibnamefont{Arkhipov}},
\bibinfo{author}{\bibfnamefont{I.}~ \bibnamefont{Babushkin}},
\bibinfo{author}{\bibfnamefont{A.~A.} \bibnamefont{Kalinichev}},
\bibinfo{author}{\bibfnamefont{A.} ~\bibnamefont{Demircan}},
\bibinfo{author}{\bibfnamefont{U.}~ \bibnamefont{Morgner}} \bibnamefont{and}
\bibinfo{author}{\bibfnamefont{N.~N.} \bibnamefont{Rosanov }},
\bibinfo{journal}{Laser Phys.lett.} \textbf{\bibinfo{volume}{15}},
\bibinfo{pages}{075003} (\bibinfo{year}{2018}).

\bibitem[{\citenamefont{Adamashvili, Knorr}(2006)}]{Adamashvili:OptLett:06}
\bibinfo{author}{\bibfnamefont{G.~T.} \bibnamefont{Adamashvili}}
  \bibnamefont{and} \bibinfo{author}{\bibfnamefont{A.}~\bibnamefont{Knorr}},
  \bibinfo{journal}{Optics Lett.},  \textbf{\bibinfo{volume}{31}},
  \bibinfo{pages}{74} (\bibinfo{year}{2006}).

\bibitem[{\citenamefont{Adamashvili }(2019)}]{Adamashvili:Phys.Rev.A:19}
\bibinfo{author}{\bibfnamefont{G.~T.} \bibnamefont{Adamashvili}} \bibnamefont{and}
\bibinfo{author}{\bibfnamefont{D.~J.} \bibnamefont{Kaup}},
\bibinfo{journal}{Phys. Phys. A.}  \textbf{\bibinfo{volume}{99}},
\bibinfo{pages}{013832} (\bibinfo{year}{2019}).

\bibitem[{\citenamefont{Diels and Hahn}(1974)}]{Diels:PhysRev:74}
\bibinfo{author}{\bibfnamefont{J.~C.} \bibnamefont{Diels}} \bibnamefont{and}
  \bibinfo{author}{\bibfnamefont{E.~L.} \bibnamefont{Hahn}},
  \bibinfo{journal}{Phys. Rev. A} \textbf{\bibinfo{volume}{10}},
  \bibinfo{pages}{2501} (\bibinfo{year}{1974}).

\bibitem[{\citenamefont{Talukder}(1969)}]{Menyuk::2010}
\bibinfo{author}{\bibfnamefont{M.~A.} \bibnamefont{Talukder}}\bibnamefont{and}
\bibinfo{author}{\bibfnamefont{C.~R.} \bibnamefont{Menyuk}},
  \bibinfo{journal}{Opt.Express} \textbf{\bibinfo{volume}{18}},
  \bibinfo{pages}{5639} (\bibinfo{year}{2010}).

\bibitem[{\citenamefont{Kivshar and Agrawal}(2003)}]{KivAg::03}
\bibinfo{author}{\bibfnamefont{Y.~S.} \bibnamefont{Kivshar}} \bibnamefont{and}
  \bibinfo{author}{\bibfnamefont{G.~P.} \bibnamefont{Agrawal}},
  \emph{\bibinfo{title}{Optical solitons. From Fibers to Photonic Crystals}}
  (\bibinfo{publisher}{Academic Press}, \bibinfo{year}{2003}).

\bibitem[{\citenamefont{Adamashvili }(2011)\citenamefont{Adamashvili}}]{Adamashvili:Result:11}
\bibinfo{author}{\bibfnamefont{G.~T.} \bibnamefont{Adamashvili}},
    \bibinfo{journal}{Results in  Physics},  \textbf{\bibinfo{volume}{1}},
  \bibinfo{pages}{26} (\bibinfo{year}{2011}).

\bibitem[{\citenamefont{Adamashvili}(2012)\citenamefont{Adamashvili }}]{Adamashvili:Optics and spectroscopy:2012}
\bibinfo{author}{\bibfnamefont{G.~T.} \bibnamefont{Adamashvili}},
\bibinfo{journal}{Optics and spectroscopy,}  \textbf{\bibinfo{volume}{113}},
\bibinfo{pages}{1} (\bibinfo{year}{2012}).

\bibitem[{\citenamefont{Adamashvili}(2009)\citenamefont{Adamashvili }}]{Adamashvili:OS:19}
\bibinfo{author}{\bibfnamefont{G.~T.} \bibnamefont{Adamashvili}},
   \bibinfo{journal}{Optics and Spectroskopy}  \textbf{\bibinfo{volume}{127}},
  \bibinfo{pages}{865} (\bibinfo{year}{2019}).

\bibitem[{\citenamefont{Adamashvili }(2012)}]{Adamashvili:PhysRevE:12}
\bibinfo{author}{\bibfnamefont{G.~T.} \bibnamefont{Adamashvili}},
  \bibinfo{journal}{Phys. Rev. E},  \textbf{\bibinfo{volume}{85}},
  \bibinfo{pages}{067601} (\bibinfo{year}{2012}).

\bibitem[{\citenamefont{Adamashvili }(2019)\citenamefont{Adamshvili,  }}]{Adamshvili:Arxiv:2019}
\bibinfo{author}{\bibfnamefont{G.~T.} \bibnamefont{Adamashvili}},
    \bibinfo{journal}{Preprint, Arxiv: 1907.10883v1,}
 (\bibinfo{year}{25 Jul 2019}).

\bibitem[{\citenamefont{Novikov et~al.}(1984)\citenamefont{Novikov}}]{Novikov::84}
\bibinfo{author}{\bibfnamefont{S.~P.} \bibnamefont{Novikov}},
\bibinfo{author}{\bibfnamefont{S.~V.} \bibnamefont{Manakov}}
\bibinfo{author}{\bibfnamefont{L.~P.} \bibnamefont{Pitaevski}} \bibnamefont{and}
\bibinfo{author}{\bibfnamefont{V.~E.} \bibnamefont{Zakharov}} ,
\bibinfo{journal}{\emph{Theory of Solitons: The Inverse Scattering Method}, (Academy of Science of the USSR, Moscow, USSR. 1984).}

\bibitem[{\citenamefont{Dodd}(2006)}]{Dodd::1982}
\bibinfo{author}{\bibfnamefont{R.~K.} \bibnamefont{Dodd}},
\bibinfo{author}{\bibfnamefont{J.~C.} \bibnamefont{Eilbeck}},
\bibinfo{author}{\bibfnamefont{J.~D.} \bibnamefont{Gibbon}} \bibnamefont{and}
\bibinfo{author}{\bibfnamefont{H.~C.} \bibnamefont{Morris}}
\bibinfo{journal}{\emph{Solitons and Nonlinear wave Equations}, Academic Press. Inc.} (\bibinfo{year}{1982}).

\bibitem[{\citenamefont{Ablowitz}(1974)}]{Ablowitz::81}
\bibinfo{author}{\bibfnamefont{M.~J.}\bibnamefont{Ablowitz}} \bibnamefont{and}
\bibinfo{author}{\bibfnamefont{H.} \bibnamefont{Segur}},
\bibinfo{journal}{\emph{Solitons and Inverse Scattering Transform}},
\bibinfo{pages}{(SIAM Philadelphia)} (\bibinfo{year}{1981}).

\bibitem[{\citenamefont{Ablowitz}(1973)}]{Ablowitz1::73}
\bibinfo{author}{\bibfnamefont{M.~J.} \bibnamefont{Ablowitz}},
\bibinfo{author}{\bibfnamefont{D.~J.} \bibnamefont{Kaup}},
\bibinfo{author}{\bibfnamefont{A.~C.} \bibnamefont{Newell}} \bibnamefont{and}
\bibinfo{author}{\bibfnamefont{H.} \bibnamefont{Segur}},
\bibinfo{journal}{ Phys. Rev. Lett.,} \textbf{\bibinfo{volume}{30}},
\bibinfo{pages}{1262} (\bibinfo{year}{1973}).

\bibitem[{\citenamefont{ Leblond}(2008)}]{Leblond::08}
\bibinfo{author}{\bibfnamefont{H.}~ \bibnamefont{Leblond}},
\bibinfo{journal}{J.Phys.B.} \textbf{\bibinfo{volume}{41}},
\bibinfo{pages}{043001} (\bibinfo{year}{2008}).

\bibitem[{\citenamefont{ Taniuti}(1973)}]{Taniuti::1973}
\bibinfo{author}{\bibfnamefont{T.}~\bibnamefont{Taniuti}} \bibnamefont{and}
\bibinfo{author}{\bibfnamefont{N.}~\bibnamefont{Iajima}},
\bibinfo{journal}{J. Math. Phys.} \textbf{\bibinfo{volume}{14}},
\bibinfo{pages}{1389} (\bibinfo{year}{1973}).

\bibitem[{\citenamefont{Adamashvili et~al.}(2014)}]{Adamashvili:Physica B:14}
\bibinfo{author}{\bibfnamefont{G.~T.} \bibnamefont{Adamashvili}},
\bibinfo{journal}{Physica B}  \textbf{\bibinfo{volume}{454}},
\bibinfo{pages}{45} (\bibinfo{year}{2014}).

\bibitem[{\citenamefont{Adamashvili }(2015)}]{Adamashvili:PhysLettA:2015}
\bibinfo{author}{\bibfnamefont{G.~T.} \bibnamefont{Adamashvili}},
  \bibinfo{journal}{Phys. Lett. A} \textbf{\bibinfo{volume}{379}},
\bibinfo{pages}{218}  (\bibinfo{year}{2015}).

\bibitem[{\citenamefont{Benjamin}(1972)}]{Benjamin::1972}
\bibinfo{author}{\bibfnamefont{T.~B.} \bibnamefont{Benjamin}},
\bibinfo{author}{\bibfnamefont{J.~L.} \bibnamefont{Bona}} \bibnamefont{and}
\bibinfo{author}{\bibfnamefont{J.~J.} \bibnamefont{Mahony}},
\bibinfo{journal}{Philos. Trans. R. Soc. London, Ser. A,} \textbf{\bibinfo{volume}{272}},
\bibinfo{pages}{47} (\bibinfo{year}{1972}).

\bibitem[{\citenamefont{Guner}(2017)}]{Guner::17}
\bibinfo{author}{\bibfnamefont{O.} \bibnamefont{Guner}},
\bibinfo{journal}{J. Ocean Eng. Sci.} \textbf{\bibinfo{volume}{2}},
\bibinfo{pages}{248} (\bibinfo{year}{2017}).

\bibitem[{\citenamefont{Manafianheris}(1990)\citenamefont{Manafianheris}}]{Manafianheris::12}
\bibinfo{author}{\bibfnamefont{J.} \bibnamefont{Manafianheris}},
 \bibinfo{journal}{World Applied Sciences Journal},
  \textbf{\bibinfo{volume}{19 (12)}}, \bibinfo{pages}{1789} (\bibinfo{year}{2012}).

\bibitem[{\citenamefont{Riskin N.M.}(1996}]{Riskin ::2010}
\bibinfo{author}{\bibfnamefont{N.~M.} \bibnamefont{Riskin }}  \bibnamefont{and}
\bibinfo{author}{\bibfnamefont{D.~I.} \bibnamefont{Trubetskov}},
\emph{\bibinfo{title}{Nonlinear Waves}}
(\bibinfo{publisher}{Nauka, M,} \bibinfo{year}{2010}).

\bibitem[{\citenamefont{Khorshidi}(2012)}]{Khorshidi::16}
\bibinfo{author}{\bibfnamefont{M.} \bibnamefont{Khorshidi}},
\bibinfo{author}{\bibfnamefont{M.} \bibnamefont{Nadjafikhah}}
\bibinfo{author}{\bibfnamefont{H.} \bibnamefont{Jafari}} \bibnamefont{and}
\bibinfo{author}{\bibfnamefont{M.} \bibnamefont{Al Qurashi}},
\bibinfo{journal}{Open Math.}, \textbf{\bibinfo{volume}{14}},
\bibinfo{pages}{1138}  (\bibinfo{year}{2016}).

\bibitem[{\citenamefont{Triki}(2012)}]{Triki::12}
\bibinfo{author}{\bibfnamefont{H.} \bibnamefont{Triki}},
\bibinfo{author}{\bibfnamefont{H.} \bibnamefont{Leblond}} \bibnamefont{and}
\bibinfo{author}{\bibfnamefont{D.} \bibnamefont{Mihalache}},
\bibinfo{journal}{Opt. Commun.} \textbf{\bibinfo{volume}{285}},
\bibinfo{pages}{3179}  (\bibinfo{year}{2012}).

\bibitem[{\citenamefont{Wazwaz}(2017)}]{Wazwaz::17}
\bibinfo{author}{\bibfnamefont{O.} \bibnamefont{Wazwaz}},
\bibinfo{journal}{J. Ocean Eng. Sci.} \textbf{\bibinfo{volume}{2}},
\bibinfo{pages}{1} (\bibinfo{year}{2017}).

\bibitem[{\citenamefont{Ghanbari}(2012)}]{Ghanbari::19}
\bibinfo{author}{\bibfnamefont{B.} \bibnamefont{Ghanbari}},
\bibinfo{author}{\bibfnamefont{D.} \bibnamefont{Baleanu}} \bibnamefont{and}
\bibinfo{author}{\bibfnamefont{M.} \bibnamefont{Al Qurashi}},
\bibinfo{journal}{Symmetry}, \textbf{\bibinfo{volume}{11}},
\bibinfo{pages}{20}  (\bibinfo{year}{2019}).

\bibitem[{\citenamefont{Khater}(2012)}]{Khater::17}
\bibinfo{author}{\bibfnamefont{M.~M.} \bibnamefont{Khater}},
\bibinfo{author}{\bibfnamefont{D.} \bibnamefont{Lu}} \bibnamefont{and}
\bibinfo{author}{\bibfnamefont{E.~H.~M.} \bibnamefont{Zahran}},
\bibinfo{journal}{Appl.Math.Inf.Sci.}, \textbf{\bibinfo{volume}{11}},
\bibinfo{pages}{1347}  (\bibinfo{year}{2017}).

\bibitem[{\citenamefont{Vinogradova}(2016)}]{Vinogradova::90}
\bibinfo{author}{\bibfnamefont{M.~B.} \bibnamefont{Vinogradova}}
\bibinfo{author}{\bibfnamefont{O.~V.} \bibnamefont{Rudenko}}
\bibnamefont{and}
\bibinfo{author}{\bibfnamefont{A.~P.}~\bibnamefont{Suhorukov}},
  (\bibinfo{publisher}{Theoria Voln. Nauka, Moscow}, \bibinfo{year}{1990}).

\bibitem[{\citenamefont{Adamashvili et~al.}(2012)}]{Adamashvili:Eur.Phys.J.D.:12}
\bibinfo{author}{\bibfnamefont{G.~T.} \bibnamefont{Adamashvili}},
\bibinfo{journal}{The Eur. Phys. J. D}  \textbf{\bibinfo{volume}{66}},
\bibinfo{pages}{101} (\bibinfo{year}{2012}).



\end{thebibliography}
\end{document}